\documentclass[11pt,letterpaper]{JHEP}
\title{Poisson Chern-Simons Gauge Theory}
\author{Bogdan Morariu\thanks{morariu@summit.rockefeller.edu} 
\\Department of Physics, Rockefeller University   \\
        New York, NY 10021}
\preprint{ RU-01-11-B}

\abstract{We formulate Poisson Chern-Simons gauge theories on compact
group manifolds. These describe a sector of the large representation
limit of noncommutative Chern-Simons in the same way as the light-cone
formulation of the membrane action describe a sector of the large $N$
Matrix model. While the formulation we give is on a group manifold,
only excitations that are invariant under the left action of the stability
group of a weight are allowed.}


\usepackage{amsmath,amssymb}

\def\11{\mbox{$1$}}

\newcommand{\rref}[1]{(\ref{#1})}
\newcommand{\beqn}{\begin{equation}}
\newcommand{\eeqn}{\end{equation}}
\newcommand{\beqarr}{\begin{eqnarray}}
\newcommand{\eeqarr}{\end{eqnarray}}

\newcommand{\matc}{\begin{array}{c}}
\newcommand{\matcc}{\begin{array}{cc}}
\newcommand{\matccc}{\begin{array}{ccc}}
\newcommand{\matcccc}{\begin{array}{cccc}}
\newcommand{\emat}{\end{array}}

\newcommand{\IH}{\relax{\rm I\kern-.18em H}}
\newcommand{\IR}{\relax{\rm I\kern-.18em R}}
\newcommand{\IK}{\relax{\rm I\kern-.18em K}}
\newcommand{\II}{\hbox{\rm 1\kern-.28em I}}
\newcommand{\Is}{\relax{\rm 1\kern-.35em 1}}

\begin{document}



\section{Introduction}
\label{Intro}
The study of noncommutative gauge theories has 
been a popular subject since their
relation to Matrix theory~\cite{Connes:1998cr} and String 
theory~\cite{Chu:1999qz,Ardalan:2000av,Schomerus:1999ug,Seiberg:1999vs}
has been understood. They are interesting because they
preserve some of the nonlocal properties inherent in String 
theory. For example, T-duality is a manifest 
symmetry~\cite{Schwarz:1998qj,Brace:1999ku}. 
For a recent  review of noncommutative gauge theory see~\cite{Douglas:2001ba}.

It is generally assumed that in the limit of a vanishing
noncommutativity parameter a noncommutative gauge theory reduces to a
commutative one. Whether this is the case depends on the
precise way the limit is taken.
The infinitesimal gauge transformations in a noncommutative gauge
theory take the form
\[
\delta \hat{A}
=
d\hat{\epsilon} - i[\hat{A}\stackrel{\star}{,} \hat{\epsilon}]~.
\]
Let us introduce the
rescaled quantities $A   =  \theta \hat{A}$\, and  
$\epsilon   =  \theta \hat{\epsilon}$\,, where $\theta$ is 
the noncommutative parameter. Then, gauge transformations written 
in terms of the rescaled quantities take the form
\[
\delta {A}
=
d{\epsilon} +\frac{1}{i\theta}[{A}\stackrel{\star}{,} {\epsilon}]~.
\]
If we take the $\theta \rightarrow 0$ limit the second term does not
vanish, rather it becomes a Poisson bracket
\begin{equation}
\delta {A}
=
d{\epsilon} + \{ {A}, {\epsilon} \}~.
\label{PoissonGT}
\end{equation}
I will refer to these type of gauge transformation as a Poisson gauge
transformation.  It has appeared in the light cone formulation of
the membrane action~\cite{Hoppe} 
and in the Lagrangian fluid dynamics
formulation of the quantum Hall 
effect\footnote{The spatial coordinate functions of the fluid are defined as 
$X^i= x^i+\theta\epsilon^{ij}\hat{A}_i$ thus they are related to $A_i$
by a translation and a rotation.}~\cite{Susskind:2001fb}.
It was also discussed in the context of noncommutative gauge theories
of D-branes~\cite{Cornalba:2000hn}.

Here we will present another example, the
Poisson Chern-Simons gauge theory on a group manifold. It can be
obtained formally as the large representation limit of the
noncommutative Chern-Simons gauge theory discussed 
in~\cite{Klimcik:1999uk,Alekseev:2000fd,Gross:2001dv}. 
However, I will introduce it independently of its noncommutative partner.
One reason to study these models is that we can understand their
geometry. This is not always possible for the noncommutative versions. 
In particular, the ``dimension''
of a noncommutative space is not a very well defined quantity; it
only makes sense in the commutative limit. In the commutative limit
for the $SU(2)$ case, we can clarify the relation between 
the fuzzy sphere of~\cite{Klimcik:1999uk,Alekseev:2000fd} and the 
the three-fold of~\cite{Gross:2001dv}. In particular we find that
while the action can be naturally written on a three dimensional
manifold, the component fields must be invariant under a $U(1)$ action
so they naturally live on a two sphere.

The plan of the paper is as follows. In Section~\ref{groupCS},
I will review the formulation of gauge theory on a group
manifold. In Section~\ref{PoissonBRST}, I describe how to introduce a
Poisson bracket on functions on the group which are invariant under
the left action of a stability subgroup. This is closely related to the
Lie-Poisson bracket on the dual of the Lie algebra of the group. I will
also extend the bracket to the exterior algebra and use it to give a BRST
formulation of the calculus. In the last section I will introduce
the Poisson Chern-Simons action, first for the $SU(2)$ case, and then
for an arbitrary compact Lie group $G$\,. 
For $G\neq SU(2)$\, I will also point out 
some difficulties with interpreting the noncommutative Chern-Simons, as 
proposed in~\cite{Gross:2001dv}, as a toy model for Witten's string field 
theory~\cite{Witten:1986cc}\,.

\section{Gauge Theory on Group Manifolds}
\label{groupCS}

In order to fix the notation and to make the paper more
self-contained,  I will give a brief review of gauge theory
on a group manifold $G$\,.
I will assume that $G$ is compact, connected and 
simply-connected. For simplicity I will also asume that $G$ is simple,
but most of what follows also works for semi-simple groups. 
Since the Lie algebra valued $1$-form 
$g^{-1}{\rm d} g$ 
is left  invariant we have the following expansion
\[
g^{-1} {\rm d} g
=T_a c^a~,
\]
where the $T_a$'s are antihermitean and satisfy the Lie algebra 
$[T_a,T_b]=f_{ab}^{~~c} \,T_c$\,, and $c^a$ are left invariant
1-forms. The forms $c^a$ give a globally defined vielbein on the group
manifold. 
The exterior derivative acting on a function $F$ on the
group can be written as
\begin{equation}
{\rm d}F=c^a\, L_a(F)~,\label{d}
\end{equation}
where $L_a$ are left invariant vector fields satisfying
$[L_a,L_b]=f_{ab}^{~~c} \,L_c$\,. Explicitly, they
are given by
\[
L_a= {\rm tr}(T_a g \frac{\partial}{\partial g^T})~.
\]
We can also extend~\rref{d} to the exterior derivative on an arbitrary
form. Let $b_a$ be an operator acting on a form $\omega$ as
\[
b_a(\omega)=i_{L_a}\omega~,
\]
where $i$ denotes the inner product of the vector field $L_a$
with the form $\omega$\,.
Then the exterior derivative can be written as
\begin{equation}
{\rm d} =c^a L_a-\frac{1}{2}f_{ab}^{~~c}\,c^a c^b b_c~,
\label{dBRST}
\end{equation}
where we assume that $L_a$ only acts on the component
functions. Note that $b_a$ and $c^a$ are odd and satisfy the ghost
anti-commutation relations $\{b_a,c^d\}
=\delta_{\,a}^{\,d}$\,.

Let us now consider gauge theory on the group manifold $G$. Later we
will mainly be interested in a $U(1)$ gauge theory but for now  we
will write our formulae so that they can also be used in the
nonabelian 
case\footnote{One should not confuse the commutators in this section 
which would not be present in an abelian theory, with the $\star$-commutators 
appearing in a noncommutative gauge theory.}.
The gauge potential $A$ can be written globally as
\begin{equation}
A=c^a A_a~,
\label{A}
\end{equation}
where we have made an implicit choice of the bundle. Using the
definition of the exterior derivative~\rref{dBRST}, one can show that the
components of the field strength $F={\rm d} A -iA^2=\frac{1}{2} c^a
c^b F_{ab}$ take the form
\[
F_{ab}=
L_a(A_b)-L_b(A_a)-i[A_a,A_b]-f_{ab}^{~~c}A_c~.
\]
Similarly, one can show that the  global and infinitesimal gauge
transformations have the form
\begin{eqnarray}
  A'_a ~&=& g^{-1}A_a g+ig^{-1}L_a (g)~,\nonumber \\
\delta A_a &=&L_a(\epsilon)-i[A_a,\epsilon]~. \label{Gtrans}
\end{eqnarray}
The Yang-Mills action is given by
\[
{\cal S}_{YM} =\int \, {\rm d}g\, \frac{1}{4} {\rm tr}\, (F_{ab}
F^{ab})~,
\]
where the tangent space indexes are raised and lowered with
$\delta_{ab}$ and $ {\rm d}g$ denotes the Haar measure on the
group. 

The Chern-Simons form $\omega_{CS}={\rm tr}(A{\rm d}A
+\frac{2}{3i}A^3)$ takes the following form
\[
\omega_{CS}=
c^a c^b c^c {\rm tr}\, 
(A_a L_b (A_c)
-\frac{1}{2}A_a f_{bc}^{~~d}A_d
+\frac{1}{3i}A_a [A_b,A_c])~.
\]
For the three-dimensional group $SU(2)$\,, we can also define a 
Chern-Simons action~\cite{Witten:1989hf}
\[
{\cal S}_{CS}=\frac{k}{4\pi}
\int \omega_{CS}~,
\]
by integrating the $3$-form $\omega_{CS}$\, over the group manifold.

\section{Poisson Algebra and BRST Calculus}
\label{PoissonBRST}
For a $U(1)$ gauge theory the gauge transformation~\rref{Gtrans} is
simply $\delta A_a =L_a(\epsilon)$\,.
We would like to modify the $U(1)$ Yang-Mills and Chern-Simons actions
above in such a way that the gauge transformation becomes similar to
the one discussed in the introduction
\begin{equation}
\delta A_a = L_a(\epsilon)+\{A_a,\epsilon\}~.
\label{PoissonGauge}
\end{equation}
First, however we must pick an appropriate Poisson bracket.
While there are known examples of Poisson brackets on groups (related
to quantum groups) we will not be interested in these. Instead I will
consider another  bracket which can only be defined on a subset of
functions on the group. Let $\lambda$ be an element of the Lie algebra
of $G$. We assume that $G$ is semi-simple so we can also think of
$\lambda$ as an element of the dual of the Lie algebra. 
We denote by $H_{\lambda}$ the stability subgroup of
$\lambda$\, i.e. $[H_{\lambda},\lambda]=0$\,. Then the bracket
can only be used on functions $F$ invariant under left multiplication
by elements $h$ of $H_{\lambda}$\,
\begin{equation}
F(hg) = F(g)~.\label{LeftInv}
\end{equation}
These functions can be identified with functions on\,
\raisebox{-.4ex}{$H_{\lambda}$}\hspace{-.03in}$\backslash G$
and this space can in turn be identified with the coadjoint orbit on
the dual of the Lie algebra of $G$ passing through $\lambda$\,. But this is a
symplectic leaf of the Lie-Poisson bracket and thus is has a natural 
Poisson bracket~\cite{BKAAK}. 
If $x_a$ denotes linear coordinates on the dual of
the Lie algebra of $G$,  the Lie-Poisson bracket on these generators 
is given by
\begin{equation}
\{x_a,x_b\}=f_{ab}^{~~c} x_c~.
\label{LiePoisson}
\end{equation}
The map $g \rightarrow g^{-1}\lambda g$ 
can be thought of as a map from\,
\raisebox{-.4ex}{$H_{\lambda}$}\hspace{-.03in}$\backslash G$ 
to the dual of the Lie
algebra. Note that two points $g$ and $g'$ belonging to the same coset of\,
\raisebox{-.4ex}{$H_{\lambda}$}\hspace{-.03in}$\backslash G$ are mapped to the 
same point. The algebra of function
on the group satisfying~\rref{LeftInv} is generated by
\[
x_a={\rm tr}(T_a g^{-1} \lambda g)~,
\]
so in principle all we need is the
bracket~\rref{LiePoisson}\,. However it is interesting to have the
bracket defined directly on the group. One can show that the bracket
takes the form
\begin{equation}
\{F,G\} ={\cal P}^{ab} \,L_a(F) \, L_b(G)~,
\label{GPoisson}
\end{equation}
where ${\cal P}^{ab}$ is defined as follows. First let $M_{ab}\equiv
f_{ab}^{~~c}x_c$\,. Note that $M$ is a degenerate antisymmetric
matrix. If fact the number of zero eigenvalues coincide with the
codimension of the symplectic leaf. Then, ${\cal P}^{ab}$ is defined
to act as {\em minus} the inverse of $M$ in the subspace where $M$ in
nondegenerate and to act as zero in the orthogonal subspace. 
For $SU(2)$ we have ${\cal P}^{ab}= (x^d x_d)^{-1}f^{abc}x_c$ and the Poisson
bracket takes the form
\begin{equation}
\{F,G\} =  \frac{1}{(x^d x_d)} f^{abc}x_a\,L_b(F) \,L_c(G)~.
\nonumber
\end{equation}
We will not need to know the explicit form of ${\cal P}^{ab}$ and will
use only the fact that it must satisfy
\begin{equation}
M_{ab}{\cal P}^{bc}M_{cd}=-M_{ad}~,
\label{MPM}
\end{equation}
which is a direct consequence of its definition. To show
that~\rref{GPoisson} gives the correct Poisson bracket it is enough to
check that on the generators $x_a$ it
gives~\rref{LiePoisson}. This can be easily accomplished using~\rref{MPM}
Note also, that when one of the entries in~\rref{GPoisson} is $x_a$\,,
we have $\{x_a,F\}=L_a(F)$\,. 

Next we extend~\rref{GPoisson} from left
$H_{\lambda}$-invariant function to left \mbox{$H_{\lambda}$-invariant}
forms. Such forms can be expanded in components as
\[
\chi=
\frac{1}{p!}\,c^{a_1}\ldots c^{a_p}\,\chi_{a_1\ldots a_p}~,
\]
where $\chi_{a_1\ldots a_p}$ are left $H_{\lambda}$-invariant
functions on the group. The graded bracket is uniquely defined by
requiring that it acts as a derivation on each of its entries, that
$c^a$ have trivial bracket with the components and satisfy
\[
\{c^a,c^b\}=0~,
\]
and that on components it reduces to the Poisson bracket~\rref{GPoisson}\,.

One can further extend the graded bracket to include the odd inner
product operators $b_a$. Then, since $\{x_a, F\}= L_a(F)$ we can
rewrite the exterior derivative acting on forms in BRST form
\[
d\chi =\{Q,\chi\}~,
\]
where the BRST function is given by
\[
Q= c^a x_a -\frac{1}{2}f_{ab}^{~~c}c^a c^b b_c~.
\]
Note that $\{Q,Q\}=0$\,, and this guarantees the nilpotency of the exterior
derivative.

\section{Poisson Chern-Simons on Compact Lie Groups}
\label{Group}
Having defined the Poisson bracket~\rref{GPoisson} and its graded
extension, we can now try to find a Poisson Chern-Simons action
invariant under the Poisson gauge transformation~\rref{PoissonGauge}.
We will do so first for $G=SU(2)$ and then generalize it to an
arbitrary group.

\subsection{The $SU(2)$ Case}
\label{SU2}
Experience with noncommutative gauge theories has taught us that
their actions look similar to nonabelian gauge theories with
$\star$-commutators instead of nonabelian commutators. Since these
would become Poisson brackets in the limit that we are considering, it
is not difficult to guess the form of the Poisson Chern-Simons
action 
\begin{equation}
{\cal S}_{PCS}=\frac{k}{4\pi}
\int A\{Q,A\}+\frac{1}{3}A\{A,A\}~.
\label{SUPCS}
\end{equation}
The integral in~\rref{SUPCS} is just the integral of a $3$-form on
the $SU(2)$ group manifold. It has two important properties. If
$\chi^p$ denotes a $p$-form the integral satisfies
\begin{eqnarray}
\int \, \{Q,\chi^2\}&=&0~, \label{IQ}\\
\int \, \{\chi^1,\chi^2\}&=&\int \, \{\chi^2,\chi^1\}=0~.
\label{CICI}
\end{eqnarray}
Since $\{Q,\chi^2\}={\rm d}\chi^2$\, 
the relation~\rref{IQ} reflects the fact that $SU(2)$ is a manifold
without boundary. I will now sketch how to prove~\rref{CICI}. First,
by writing the forms in components, one can show that~\rref{CICI} 
is a consequence of the following relation
\begin{equation}
\int\, {\rm d}g \,\{F,G\}=0~,
\label{IFG}
\end{equation}
where $F$ and $G$ are arbitrary left
$H_{\lambda}$-invariant functions in the group. We have
\begin{eqnarray}
\int {\rm d} g\, \{ F,G\}~~~~=
~~~~\int {\rm d} g\, \{ F,x_a\}\partial_a G~~~~~~~~~~~~~~~~&=&
\nonumber \\
\int {\rm d} g\,[\{F\partial_a G,x_a
\}-F\{\partial_aG,x_a\}]~~~~~~&=&
\nonumber\\
\int {\rm d} g\,[-L_a(F\partial_a G)+ F\partial_a\partial_b G
\{x_a,x_b\}]&=&0~.\nonumber
\end{eqnarray}
The last equality follows from the invariance of the Haar measure
$\int\, {\rm d}g \,L_a(F)=0$\, and the antisymmetry of the
Poisson bracket. Note that~\rref{IFG} is true for any compact Lie
group not just for $SU(2)$\,.
Using~\rref{IQ} and~\rref{CICI} it is not difficult 
to prove the gauge invariance of the action (the Poisson Chern-Simons form  
$\omega_{PCS}=A\{Q,A\}+\frac{1}{3}A\{A,A\}$\, is not invariant under
this gauge transformation, only the action is).

The equations of motion derived from~\rref{SUPCS} can be written compactly 
\begin{equation}
\{Q+A,Q+A\}=0~.\label{EOM}
\end{equation}
Note also that the gauge transformation~\rref{PoissonGauge} can be written as
\begin{equation}
\delta A=\{Q+A,\epsilon\}~.
\label{Qgauge}
\end{equation}
To obtain the spectrum, it is enough to consider the 
linearized equations of motion~\rref{EOM} and
gauge transformations~\rref{Qgauge}, which take the form
\begin{equation}
\{Q,A\}=0~,~~~\delta A=\{Q,\epsilon\}~.
\end{equation}
To this order, the solutions of the equations of motion modulo
gauge transformations coincide with the cohomology of $Q$\,.

In components the Poisson Chern-Simons action takes the form
\begin{equation}
{\cal S}_{PCS}=\frac{k}{4\pi}
\int \,{\rm d}g\,\epsilon^{abc}\,
(A_a \{x_b, A_c\}
-\frac{1}{2}A_a f_{bc}^{~~d}A_d
+\frac{1}{3}A_a \{A_b,A_c\})~,\label{Scomp}
\end{equation}
while the infinitesimal Poisson gauge transformation is given by 
\begin{equation}
\delta A_a =L_a(\epsilon)+\{A_a,\epsilon\}=\{x_a+A_a,\epsilon\}~.
\label{PGT}
\end{equation}
The equations of motion, written in components, have the form
$F_{ab}=0$\,, with the field strength $F_{ab}$ given by
\begin{equation}
F_{ab}=
L_a(A_b)-L_b(A_a)+\{A_a,A_b\}-f_{ab}^{~~c}A_c
=\{x_a+A_a,x_b+A_b\}-f_{ab}^{~~c}(x_c+A_c)~.
\label{FS}
\end{equation}

We have normalized the Poisson Chern-Simons action in a similar way as
a standard Chern-Simons action. However the level $k$ does not seem to
be quantized\footnote{This is in contract to Chern-Simons on the 
noncommutative plane where the level is 
quantized~\cite{Nair:2001rt,Bak:2001ze}.}. 
In the noncommutative Chern-Simons 
theory~\cite{Klimcik:1999uk,Alekseev:2000fd,Gross:2001dv} one can
explicitly check the invariance of the action under global gauge
transformations. It should also be possible to do this directly in the
Poisson Chern-Simons case, but we will not do this here. Even
if we generalize the theory in the obvious way to a $U(n)$ Poisson
Chern-Simons the level is not quantized. However, note
that the lack of quantization of the level is not so
surprising. Indeed, if one scales out the Poisson bracket, the action
reduces to a standard $U(n)$ Chern-Simons but we still have the
invariance under $H_{\lambda}=U(1)$. Thus instead of $\pi_3(U(n))$\,, 
gauge transformations are
classified by $\pi_2(U(n))$ which is trivial.

\subsection{Arbitrary Compact Lie Groups Case}
\label{LieG}
It was noticed in~\cite{Gross:2001dv} that the noncommutative
Chern-Simons gauge theory can be generalized to an arbitrary compact
group $G$. In our case we can arrive at a similar result as
follows. In equation~\rref{Scomp} we can substitute the structure
constants of $G$\,, (with appropriately placed indices,) for 
the $\epsilon$-tensor
\begin{equation}
{\cal S}_{PCS} \propto
\int \,{\rm d}g\,f^{abc}\,
(A_a L_b (A_c)
-\frac{1}{2}A_a f_{bc}^{~~d}A_d
+\frac{1}{3}A_a \{A_b,A_c\})~.\label{ScompG}
\end{equation}
One can check that this action is also invariant under the
Poisson gauge transformation~\rref{PGT}. 

We can write the
action~\rref{ScompG} in a more geometric way as follows. On any group
manifold we can define the left invariant $3$--form 
\[
\Omega={\rm tr}\,[(g^{-1} {\rm d}g )^{\wedge 3}]~.
\]
Then the action can be written 
\begin{equation}
{\cal S}_{PCS} \propto
\int \star \,\Omega \wedge \omega_{PCS}=
\int \,{\rm d}g\, <\Omega,\omega_{PCS}>~,
\label{PCSG}
\end{equation}
where $\star$ denotes the Hodge dual and 
$<\,,>$ denotes the inner product of $3$-forms.

However, it is clear that in the
action~\rref{Scomp} the $\epsilon$-tensor 
comes from the volume of integration and
it is an accident (in the three dimensional case) that it coincides with
the structure constants of the group $SU(2)$. Thus one wonders if
substituting the structure constants for the $\epsilon$-tensor does
not have some undesirable consequences. Indeed, while it is true that
the action~\rref{ScompG} is gauge invariant, the equations of motion
derived from it read
\begin{equation}
f^{abc}F_{bc}=0~,\label{fF}
\end{equation}   
and these only imply the vanishing of the field strength if the rank
of the group is three. Thus for arbitrary $G$ 
we do not have $\{Q+A,Q+A\}=0$\,. 
Similarly, in the noncommutative Chern-Simons case the equation
$(Q+A)^2=0$\, claimed in~\cite{Gross:2001dv} is only valid for $G=SU(2)$\,. 
In Witten's string field theory~\cite{Witten:1986cc} 
it is implicitly
assumed that the ``integral'' is nondegenerate, such that the equations
of motion are exactly the vanishing of the field strength. 
Therefore we can
only interpret the three dimensional case as a toy model for Witten's
string field theory. It is however an interesting problem to find
nontrivial solutions of nonvanishing field strength of the equations~\rref{fF}.

It is also possible to define a Poisson Yang-Mills gauge theory on a
group manifold  
\[
{\cal S}_{PYM} =\int \, {\rm d}g\, \frac{1}{4} \, F_{ab}
F^{ab}~,
\]
where now the field strength is given by~\rref{FS}. Note that the
Lagrangian is only covariant under Poisson gauge transformations and
just as in noncommutative gauge theories one has to integrate to obtain
a gauge invariant action.

We conclude by
noting that Poisson gauge theories are interesting because they
retain some of the properties of noncommutative gauge theories. For
example there are no local Poisson gauge invariant quantities. It
would be desirable to develop a better understanding of 
the observables of these theories. Note also that in a manner similar
to the ``regularization'' of the membrane action discussed 
in~\cite{Hoppe}, we can quantize the Poisson bracket on the
coadjoint orbits~\cite{BKAAK}, (see 
also~\cite{Alexanian:2001qj,Morariu:2001qa} for a more recent
treatment in the spirit of this paper,) and obtain the noncommutative 
Chern-Simons action. The quantization leads to unitary representations
only if $\lambda$ is a weight. In this case we obtain the
noncommutative Chern-Simons theory corresponding to the irreducible
representation whose highest weight is $\lambda$\,. It would be
interesting to find the Poisson Chern-Simons theory corresponding to
the reducible representations discussed in~\cite{Gross:2001dv}.

\acknowledgments
I would like to thank Klaus~Bering and Alexios~Polychronakos
for useful      discussions.
This work was supported in part by the U.S.~Department of Energy
under Contract Number DE-FG02-91ER40651-TASK B.

\end{document}